\documentstyle[aps,prl,epsfig]{revtex}
\begin{document}               
\tightenlines
\title{Amplitude of Aharonov-Bohm oscillations in mesoscopic metallic rings as a function of the DC bias voltage}
\author{C. Terrier $^1$, C. Strunk $^1$, T. Nussbaumer $^1$, D. Babi\'{c} $^{1,2}$ and C. Sch\"{o}nenberger $^1$}
\address{$^1$Institute of Physics, University of Basel, Klingelbergstrasse 82, CH-4056 Basel, Switzerland \\
$^2$Department of Physics, Faculty of Science, University of Zagreb, 
Bijeni\v{c}ka 32, HR-10001 Zagreb, Croatia}
\maketitle
\begin{abstract}

We report measurements of the amplitude of the Aharonov-Bohm oscillations in a mesoscopic diffusive gold ring as a function of the DC bias voltage $V_{DC}$. The amplitude of the $h/e$ oscillations increases with $V_{DC}$ once the 
Thouless energy $E_c$ and thermal energy are exceeded, and decreases at higher values of $V_{DC}$.
The increase of the amplitude is interpreted in terms of a superposition of the statistically independent contributions of $eV_{DC}/E_c$ energy intervals, whereas its decrease at high $V_{DC}$ could be attributed to enhanced inelastic scattering processes. \\

\flushleft{PACS: 73.23--b, 03.65.Bz \\ Keywords: Mesoscopic systems; Aharonov-Bohm oscillations; Ensemble averaging}
\end{abstract}
\section{Introduction}               

Direct measurability of the quantum coherence (QC) of electrons is  one of the most important characteristics of mesoscopic systems.  At these dimensions, typically 10 nm - 10 $\mu$m, there are usually still enough atoms for using statistical concepts of macroscopic solid-state objects but manifestations of quantum-mechanical phenomena are much stronger than in macroscopic systems. QC of electrons in a diffusive metal persists over a length scale $L_\phi$
typically of order of 1 $\mu$m at $T \sim 1$ K, which means that for these materials one needs samples of comparable dimensions to probe it directly.
By using electron-beam lithography (EBL),  primary fabrication method of mesoscopic physics, structuring  samples in geometries and sizes suitable for exploring various manifestations of QC becomes possible. One of the most direct observations of the coherent nature of electrons in conductors are the Aharonov-Bohm (AB) oscillations of electrical resistance in loop-shaped samples, as a function of a magnetic flux $\Phi$ threading the loop \cite{ab,alt1}.
This behaviour has been observed in thin-walled metallic cylinders of micrometre size \cite{ss,arsh,gijs}, mesoscopic rings fabricated by EBL \cite{pann,webb1,chand}, and recently even in carbon nanotubes \cite{adrian}, i.e. conducting molecules of appropriate cylindrical geometry.

The AB phenomenon originates in a fundamental property that a quantum particle has several paths available (in the case of a loop the two branches) for passing through a conductor, which results in their interference.  An applied magnetic field $\vec{B}$ introduces an additional term in the electron's phase $\varphi$, 
i.e. $(e /  \hbar) \int \vec{A} \cdot {\rm d} \vec{l}$, where $\vec{A}$ is the vector potential and the integration is carried out along the electron path.  It is easy to show that for a loop one then expects oscillations of the magnetoresistance $R$.
More precisely, $R(B)$ can be expanded in to a Fourier series with the periods $(h/e) / n$, $n \geq 1$, of a magnetic flux through the loop. At low temperatures  \cite{stone}
\begin{equation}
\Delta R(B,T) = R(B,T)-R(0,T) =\sum _{n=1}^\infty  r_n(B,\epsilon) \cos \left( 2 \pi \frac{n \Phi}{h/e} +
 \varphi_n (B,\epsilon) \right) \; ,
\end{equation}
where $\epsilon$ the electron energy relative to the Fermi energy, $r_n(B,\epsilon)$ the Fourier amplitude of the
$n$th mode and $\varphi_n (B,\epsilon)$ a phase shift.  The field dependences of $r_n$ and $\varphi_n$ 
are weak in a ring with a good aspect ratio, i.e. when the area of the hole is considerably larger than that of the 
annulus \cite{stone}. Namely, the variations of $r_n$ and $\varphi_n$ with $B$ originate from the beats caused by semiclassical trajectories enclosing slightly different areas, and for a ring with the  linewidth small in comparison with the diameter they are slow  with respect to $h/e$. Furthermore, in a typical mesoscopic ring  the length $L$ of one arm is comparable to $L_\phi$, and the $n=1$ term dominates, 
describing the simplest possible passing of an electron through the ring, i.e. with no backscattering which would give rise to the higher harmonics. This is a direct consequence of the length scales involved, as the amplitude of the $n$th term in Eq.1 is roughly proportional to $\exp(-2nL/L_\phi)$ \cite{datta}. 

In equilibrium the range of energies which contribute to the interference pattern is set by $k_B T$. If $k_B T$ is larger than
the Thouless energy $E_c \approx \hbar D / L^2$ ($D$ is the diffusion constant), this energy range subdivides in to $k_B T / E_c$ 
statistically independent energy intervals, which leads to damping of the AB amplitude by ensemble averaging. The experiment discussed in this paper is performed in the regime $k_B T > E_c$. If a  DC voltage $V_{DC}$ or a low-frequency AC voltage $V_{AC}$  (in both cases larger than $k_B T/e$ and $E_c/e$) is applied one would expect that the energy interval contributing to the ensemble averaging widens up, therefore further reducing the amplitude of the AB oscillations \cite{webb2}.  Surprisingly, in differential-conductance measurements where the signal is probed by a small AC voltage $V_{AC}$   
one observes {\it increase} of the amplitude with increasing $V_{DC} >> (V_{AC}, k_B T / e, E_c/e)$ instead, as shown recently by H\"{a}ussler et. al. for the first time \cite{hauss}. 
In this paper we confirm their finding and show that it extends to approximately three times higher temperatures (270 mK).
We discuss a possible origin of this phenomenon in more detail as well.

\section{Experimental}

The sample, a mesoscopic gold ring 1 $\mu$m in diameter (taken as the average of the inner and the outer diameters), 
20 nm thick and of the average width of about 90 nm, was realised by EBL and evaporation of the metal (see Fig.1).
With $D=116$ cm$^2$s found for our sample this gives $E_c \approx 3$ $\mu$eV.
The sample has a good aspect ratio, which simplifies interpretation of the results, as discussed
above.
We first briefly describe the fabrication of the sample. On top of a previously cleaned Si / 400 nm SiO$_2$ wafer a 
double-layer resist consisting of polymethyl methacrylate co-methacrylic acid (PMMA-MA, 450 nm) and less exposure sensitive polymethyl methacrylate (PMMA, 270 nm)  was spun and baked. The structuring of the resist by EBL was done in a JEOL JSM IC 848 SEM at 20 kV acceleration voltage and using three different current values (34 pA, 87 pA and 12 nA). The smallest value was used to create the small ring structure, suitable for four-point measurements, and the other two current values to make the contact pads. The area dose was 
450 $\mu$C/cm$^2$ and the line dose 3 nC/cm. The development was done using a 1:3 mixture of 4-methyl-2-pentanone isobutyl-methylketon and 2-propanol.

\begin{figure}
\centerline{{\psfig{file=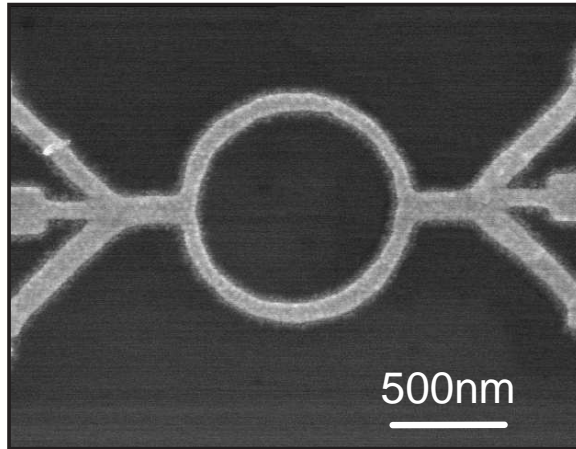,width=7.7cm}}}
\caption{SEM photograph of the sample prepared for four-point measurements. The average diameter of the ring is 
1 $\mu$m, its thickness is 20 nm and the width of the wires 90 nm in average.} 
\end{figure}

Gold was evaporated in a high-vacuum chamber at a pressure of $1\cdot 10^{-5}$ mbar. In order to avoid heating of the resists, the thermal conductivity between the sample and its holder was enhanced using a thermally-conducting paste. Adhesion of the metal to the wafer was improved by in-situ Ar-sputtering prior to the evaporation. The evaporation was carried out using three different angles (with thickness of 20 nm in the direction perpendicular to the surface of the sample and 30 nm at the inclinations of $\pm45^0$). The double-layer technique with the upper layer having a lower exposure sensitivity offers the possibility of evaporating under such large angles because it provides a sufficient resist undercut.  Thereby the contact pads can be thickened without disturbing the small ring structure \cite{henny}. This enables to fabricate the sample in only one lithography and evaporation step. The sample was further glued on to a chip carrier and the electrical connections were made by  ultrasonic bonding of 50 $\mu$m Al wires. 

The measurements were carried out at 270 mK in a $^3$He cryostat. The magnetoresistance was measured by a 
lock-in technique of low-frequency (17 Hz) and low-excitation ($eV_{AC} < k_B T$), using an Adler-Jackson bridge. We achieved a voltage sensitivity of $\sim 0.3$ nV.  
In order to investigate the influence of $V_{DC}$ on the AB effect we superposed a DC current $I_{DC}$ to the AC measurement current, which resulted in a DC bias voltage $V_{DC} = I_{DC} R_0$. $R_0= 24.6$ $\Omega$ is the sample resistance at $B=0$, corresponding to a sheet resistance $R_\Box= 1.8$ $\Omega / \Box$. 
The magnetic field sensitivity was around 0.02 mT and the step between each measurement 
point was chosen to be 0.17 mT. Fitting the expression for the 1D weak-localisation magnetoresistance at low magnetic fields \cite{alt} to our data gave estimates $L_\phi= 1.6$ $\mu$m and $\tau_\phi = L_\phi^2 /D = 1.4$ ns. At this temperature (270 mK) electron-electron scattering dominates $L_\phi$, whereas at higher temperatures (typically above $\sim$ 1 K) electron-phonon scattering becomes important as well.

\section{Results and Discussion}

As mentioned in the introduction,  new features in QC of electrons are detectable by applying a bias voltage $V_{DC}$ larger than
$E_c / e$ and $k_B T /e$ across a mesoscopic ring in an AB experiment performed using a small AC measurement current. Discussion of a possible origin of the observed {\it increase} of the amplitude of the AB oscillations with $V_{DC}$, instead of
its {\it decrease} expected from simple ensemble-averaging arguments, is central to this section.

In Fig.2 we show parts of two typical measured field-sweep traces,  for zero and  finite $V_{DC}$. The measured periodicity of 5.8 mT corresponds to $h/e$, as expected.  Contrary to the common understanding of ensemble averaging, when a DC bias voltage of 0.49 mV is applied the amplitude of the oscillations increases by a factor of 5 compared to that at $V_{DC}=0$. To analyse the data more quantitatively we carried out the Fourier analysis for each trace measured (not shown). In Fig.3 we plot the height of the $n=1$ peak of the Fourier transform versus $V_{DC}$ for two different magnetic-field intervals. A rapid increase of the peak height is observed for $V_{DC} < 0.5$ mV, which is followed by a gradual decay to small values for $V_{DC} > 0.5$ mV. Note that the applied voltages are much larger than 
$E_c/e \approx$ 3 $\mu$V and $k_B T /e= 25$ $\mu$V. The largest voltage in fact corresponds to as much as 30 K.
The oscillation amplitude around 7.4 T is enhanced with respect to that at $\sim 2$ T, which might be due to the scattering at residual magnetic impurities being suppressed in high magnetic fields.

\begin{figure}[h]
\centerline{{\psfig{file=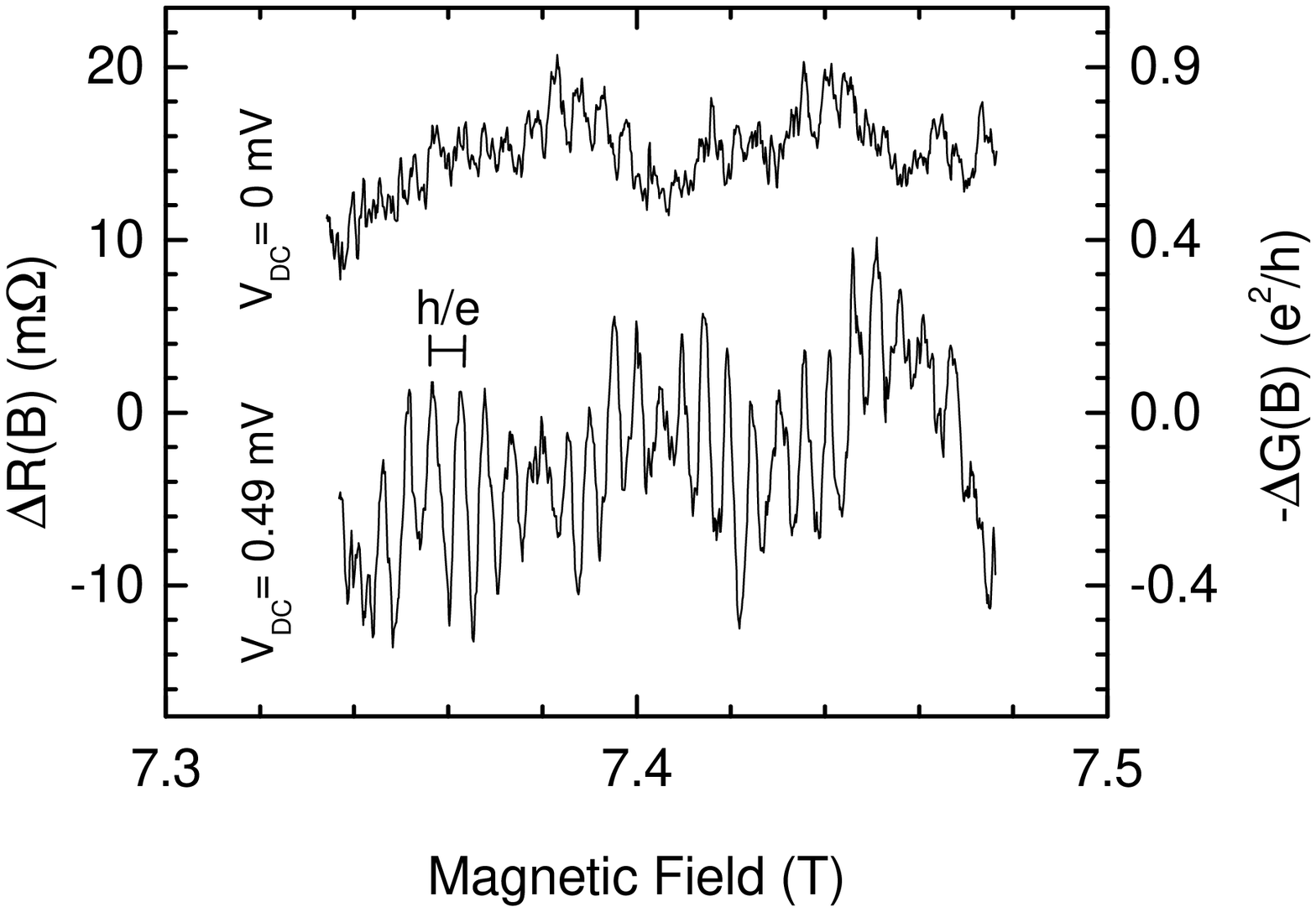,width=9cm}}}
\caption{Typical magnetoresistance traces of $h/e$ oscillations at 270 mK, for $V_{DC}=0$ and $V_{DC}=0.49$ mV.
Note the pronounced increase of the oscillation amplitude with $V_{DC}$. }
\end{figure}

As discussed by Larkin and Khmel'nitskii \cite{khm}, if a voltage considerably larger than $E_c / e$ and $k_B T /e$ is applied, the energy interval contributing to the current increases and is subdivided in to $N=eV/E_c$ statistically independent subintervals. In this case the average current $<I>$ is proportional to $V$ and its fluctuation is given by
$\Delta I \sim (e^2/ \hbar) (V E_c /e)^{1/2}$. The current fluctuations have a significant effect on the differential conductance, as follows.  The total current $I$ can be written as 
$I=<I>+ \Delta I \exp(-L/L_\phi) \delta \alpha$, where $\delta \alpha$ is a random function with the root-mean-square (rms)
amplitude unity and
variations on the voltage scale with a typical frequency $2 \pi e / E_c$. The differential conductance is in this case
\begin{equation}
\frac{{\rm d} I}{{\rm d} V} = <G> + \left( \frac{{\rm d} \Delta I}{{\rm d} V} \delta \alpha +
\Delta I \frac{{\rm d} \delta \alpha}{{\rm d} V} \right) e^{-\frac{L}{L_\phi}}
\; ,
\end{equation}
where $<G>= {\rm d} <I> / {\rm d} V$ is the mean conductance.  By assuming that the dominant variation of $\delta \alpha$
occurs on a typical scale $E_c /e$ one can approximate 
${\rm d} \delta \alpha / {\rm d} V$ by  $e \delta \alpha / E_c$. Since the first term in the brackets on the right-hand side of Eq.2
gives no contribution to the differential-conductance fluctuations for $V >> E_c/e$, we obtain

\begin{equation}
\Delta \left( \frac{{\rm d} I}{{\rm d} V} \right) \propto  \frac{e^2}{h}  \sqrt{ \frac{eV}{E_c}}  \delta \alpha 
\; e^{-\frac{L}{L_\phi}}\; .
\end{equation}

By setting $V=V_{DC}$ and taking the rms value of $\Delta ({\rm d}  I / {\rm d} V)$ we obtain 
$[\Delta ({\rm d}  I / {\rm d} V)]_{rms} \propto V_{DC}^{1/2}$. The same voltage dependance holds for the fluctuations of the differential resistance.  In order to observe an enhancement in the  fluctuations it is crucial to increase $N$ by a large $V_{DC}$ but to probe the differential conductance by a small $V_{AC}$.  In a usual conductance measurement, either DC or low-frequency AC, the current fluctuations $\Delta I \propto \sqrt{N}$, whereas the applied voltage $V \propto N$. Thus, the conductance fluctuations 
$\Delta I / V \propto 1/\sqrt{N}$, which {\it decreases} with increasing $V$. In contrast, in a non-equilibrium experiment with 
$eV_{DC} >> k_B T > e V_{AC}$ the differential conductance measures $\Delta I_{DC} / V_{AC}$, which
{\it increases} with increasing $V_{DC}$ as $\sqrt{N}$.

\begin{figure}[h]
\centerline{{\psfig{file=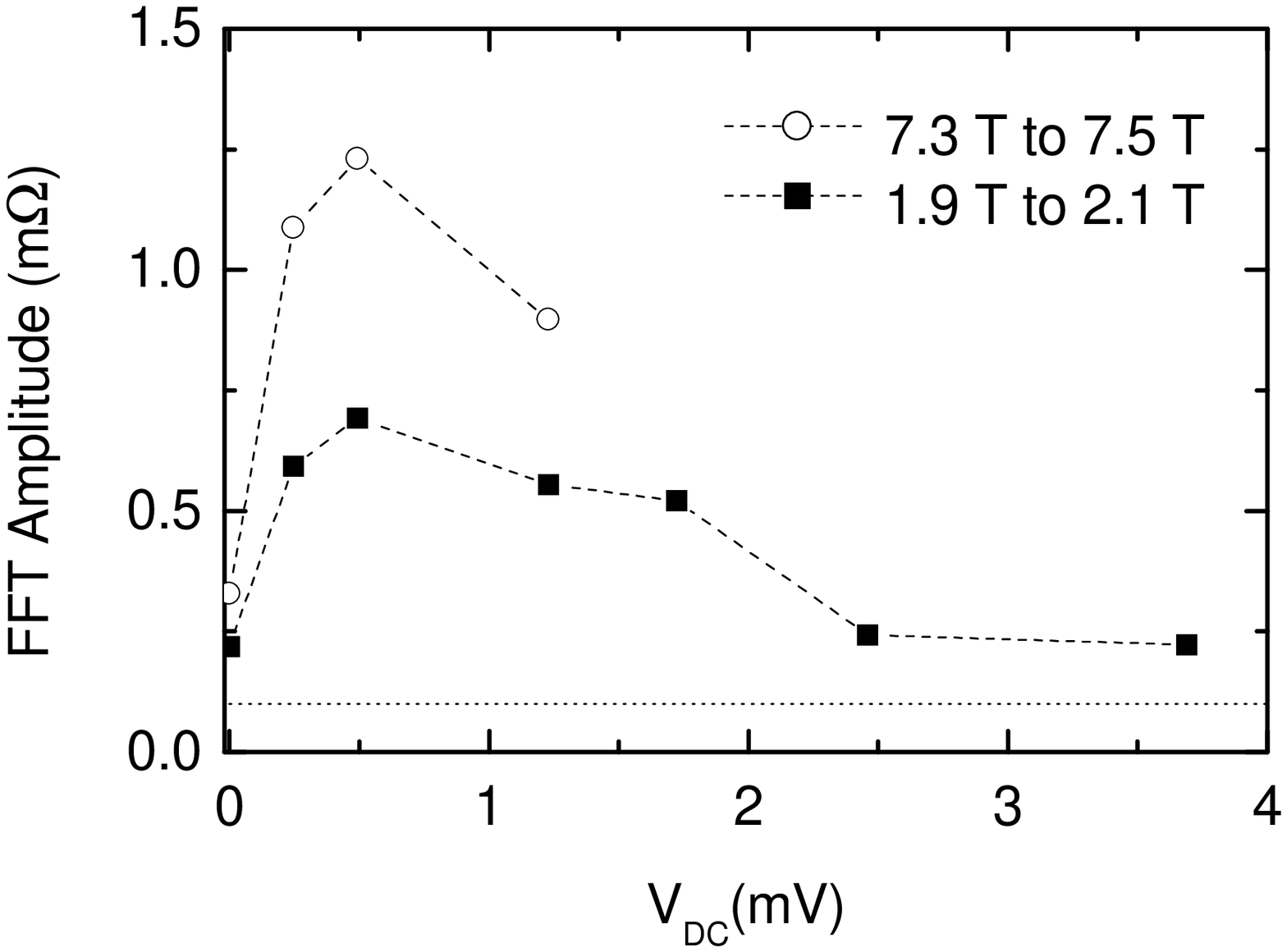,width=9cm}}}
\caption{Fourier amplitude of the $h/e$ oscillations plotted against $V_{DC}$, for two different magnetic-field intervals. The Thouless energy $E_c$ is 3 $\mu$eV, and for all the points $eV_{DC} >> E_c, k_B T$. 
The dashed line represents the noise floor of our measurements.
The amplitude at $\sim 7.4$ T is higher than that at $\sim 2$ T probably because scattering at residual magnetic impurities is suppressed in large magnetic fields.}
\end{figure}

A corresponding increase of the oscillation amplitude can be seen in Fig.3 for $V_{DC} < 0.5$ mV. The subsequent decrease of the amplitude is most probably caused by a reduction of $L_\phi$ for the electrons at high energies. These 
high-energy electrons at least partly thermalise via electron-electron collisions, which leads to an effective electron 
temperature $T_{el}$ higher than the bath temperature $T$. From a simple model calculation assuming that $T_{el}$ is governed by the diffusion cooling of electrons \cite{henny} we estimate $k_B T_{el} \approx e V_{DC}/5$.
This means that the ensemble averaging by the electron-heating effect always remains substantially smaller than
the applied DC bias voltage and explains the persistence of the AB oscillations up to moderately high $V_{DC}$.
At the maximum of the AB oscillation amplitude vs. $V_{DC}$ plot we estimate $T_{el} \sim 1.2$ K. At larger $V_{DC}$
the temperature dependence of $L_\phi$  provides  exponential reduction of the oscillation amplitude (Eq.3),
which cuts off its initial $\sqrt{V_{DC}}$ - like increase.

\section{Conclusions} 

We studied the influence of the DC bias voltage $V_{DC}$ on the amplitude of the Aharonov-Bohm oscillations in a diffusive gold ring.  If the bias voltage exceeds the Thouless energy $E_c$ (in our case 3 $\mu$eV) and thermal energy, the amplitude of the $h/e$ oscillations first increases.
We interpret this phenomenon in terms of a subdivision of the energy interval contributing to the interference in to $eV_{DC}/E_c$ statistically independent subintervals.  The interference pattern is formed as a superposition of the contributions of these subintervals. After the amplitude reaches its maximum value it decreases and is suppressed at high voltages probably because of inelastic scattering processes. Our work confirms the recent finding of H\"{a}ussler et. al. \cite{hauss} and shows the existence of the mentioned behaviour at about three times higher temperatures (270 mK). The observation of this novel phenomenon allows to study the coherent behaviour of electrons out of equilibrium.

{\bf Acknowledgements} \\
We are grateful to R. H\"{a}ussler, M. Buitelaar, F. Dewarrat, H.-W. Fink, T. Hoss, M. Kr\"{u}ger and S. Oberholzer for useful discussions. This work was supported by the Swiss National Science Foundation.

\end{document}